\documentstyle[11pt,aasms4,psfig]{article}
\def\sn{SN~1987A}
\def\s1{Source One}
\def\km/s{km~s$^{-1}$}
\def\HII{H\,{\sc ii}}

\begin{document}
\title{The asymmetric radio remnant of SN 1987A}
\author{B. M. Gaensler}
\affil{Australia Telescope National Facility, CSIRO, PO Box 76,
Epping, NSW 2121, Australia and Astrophysics Department, School of
Physics A29, University of Sydney, NSW 2006, Australia. Email: 
b.gaensler@physics.usyd.edu.au}
\and
\author{R. N. Manchester, L. Staveley-Smith,
A. K. Tzioumis, J. E. Reynolds and M. J. Kesteven}
\affil{Australia Telescope National Facility, CSIRO, PO Box 76,
Epping, NSW 2121, Australia}

\vspace{1cm}
\begin{center}
\today
\vspace{3cm}

To appear in volume 479 of the {\em The Astrophysical Journal} (20 Apr 1997).
\vspace{1cm}
\end{center}

\begin{abstract}

We present seven years of radio observations of \sn\ made with the
Australia Telescope Compact Array. At 1.4, 2.4, 4.8 and 8.6~GHz, the
flux density of the radio remnant has increased monotonically since emission
was redetected 1200 days after the explosion.  On day 3200, the remnant
was expanding at $2800\pm400$ \km/s, which we interpret as indicating
significant deceleration of the fastest moving ejecta.  Since day 1787
the spectral index has remained constant at $\alpha = -0.95\pm0.04$ ($S
\propto \nu^{\alpha}$).  These observations are all consistent with the
shock having encountered a denser, shocked, component of the
progenitor's stellar wind. At the current rate of expansion, the shock
is expected to encounter the inner optical ring in the year 2006$\pm$3,
in line with predictions made by hydrodynamic simulations.

Using super-resolution, we have also obtained 9~GHz images of the
remnant (resolution $\approx 0''.5$) at four epochs. The emission is
distributed around the rim of a near-circular shell, but has become
increasingly asymmetric with time. There are two ``hotspots'' to the
east and west, aligned along the major axis of the optical ring.  This
morphology is most likely indicative of an axisymmetric
circumstellar medium into which the shock is expanding, consistent with
present understanding of the progenitor star and its environment.
The two hotspots are increasing in flux density at different rates, which
may indicate directional anisotropies in the ejecta.  
We believe that the northern and southern regions of the
remnant are encountering a shocked wind which is less dense and also further
from the progenitor star than that in the bright regions of emission.
As a result, these regions should eventually brighten and/or extend.

\end{abstract}


\section{Introduction}

\sn\ in the Large Magellanic Cloud has presented a unique opportunity.
The nearest ($\sim$50 kpc) supernova (SN) to have been detected since
the invention of the telescope, its aftermath presents a detailed
picture of its progenitor's circumstellar medium (CSM) and of the
interaction of this CSM with the SN shock.

Since the discovery of radio supernovae (RSNe) over 25 yr ago
(\cite{gbb+72}), such objects have proven to be useful tools in probing the 
CSM of massive stars. The flux density from a typical RSN takes
weeks to years to peak in its radio emission, then undergoes a slow 
power-law decay extending over a period of years or even decades
(\cite{wvsp96}). This is representative of the SN shock sweeping up the
dense, slow-moving wind generated by the red supergiant (RSG)
progenitor (\cite{che82}).

However, \sn's progenitor, Sk--69$^\circ$202, was very different.  It is
believed to have evolved from a RSG into a {\em blue} supergiant (BSG)
approximately $2 \times 10^4$ yr before the SN event (\cite{ch91}).
During the star's period as a BSG, the high-velocity
BSG wind interacted with the slower RSG wind surrounding it, creating a
cavity.  Ground-based and HST optical observations of \sn\ have
revealed a complex nebulosity surrounding the explosion, consisting of
a central circular ring of diameter $\approx 1''.6$ (0.4~pc), and two
similar rings of diameter $\approx 3''.6$ (0.9~pc), one on either
side. The whole system is inclined at $\sim 45^{\circ}$ to the line of
sight (\cite{jab+91}; \cite{plck95}). While the exact origin for this ``triple
ring nebula'' is unclear, it appears to demarcate the interface
between the BSG and RSG winds, material at the boundary being excited
by the ultraviolet flash from the supernova (\cite{lm91b}).
The optical emission appears to indicate that
the BSG wind bubble is bipolar in shape (\cite{ckh95}), the
``standard'' model being that this is caused by an RSG wind which is
denser at the equator than at the poles (\cite{lm91b}; \cite{ma95}). 
Problems exist with this model however (\cite{blo94}; \cite{bkh+95}), 
and many alternatives have been proposed 
(e.g. \cite{ml94}; \cite{bk95a}).

Radio emission from \sn\ was detected at 0.843~GHz by the Molonglo
Observatory Synthesis Telescope (MOST) on 1987 February 25, two days
after the SN event.  This emission reached a maximum four days after
the explosion (\cite{tcb+87}), then followed a power-law decay, to become
undetectable less than a year later (\cite{bcct95}). This radio outburst
is understood to be a consequence of the rapidly moving, low-density
BSG wind, which produced only a short-lived period of
radio emission when hit by the SN shock (\cite{sm87}; \cite{cf87}).

After remaining quiescent for several years, radio emission from the
remnant of \sn\ was redetected by both the MOST and the Australia
Telescope Compact Array (ATCA) in mid-1990 (\cite{smk+92}, hereafter
Paper I), approximately
1200~d after the explosion. At 9~GHz, the ATCA's highest 
frequency, the remnant was just resolved,
and by early 1992 the emission was found to deviate
by $\sim$10\% from spherical symmetry (\cite{smk+93}, hereafter Paper~II).
Super-resolution
of ATCA data at comparatively low signal-to-noise produced images 
of resolution $\approx 0''.5$ (\cite{sbr+93}, hereafter Paper~III). 
In an image corresponding to
observations made in 1993 January, radio emission is distributed around
the rim of a shell, with up to 50\% of the flux density originating from
two bright limbs on the shell's eastern and western edges. To account
for the size of the remnant at the second radio turn-on, a velocity 
of $\sim$35\,000 \km/s was required for the fastest ejecta. However, only a
marginal increase in size was subsequently detected over a 20-month period, 
implying that the remnant had either decelerated significantly,
or was rapidly changing shape. This
emission, which has since been shown to be centred on the position of
the SN (\cite{rjs+95}), was still increasing in its intensity four years
after its reappearance (\cite{bcct95}; \cite{smt+96}, hereafter Paper~IV).

The second onset of radio emission has been interpreted as resulting from 
the shock encountering a sudden increase in the density of the CSM.  The
diameter of the emitting region is noticeably smaller than the inner
optical ring, and so the shock is yet to reach the undisturbed RSG
wind.  Chevalier (1992a)\nocite{che92b} 
has argued that the shock has reached the
termination shock of the BSG wind, so that it is now progressing through a
region of constant density. Such a theory is supported by more
detailed modelling of the radio emission
(\cite{bk92b}; \cite{dbk95}, hereafter DBK).
Chevalier \& Dwarkadas (1995\nocite{cd95},
hereafter CD95), have recently proposed an
alternative model, that the dense region that the shock has encountered
is actually an ionized component of the {\em RSG} wind located inside
the optical ring. The greater density of this RSG component can better
explain the appearance of X--ray emission from \sn\ (\cite{bbp94};
\cite{ght94}), and also the large
apparent deceleration of the SN shock suggested by the results of
Paper III.

Only two other remnants of supernovae (SNe) as young as \sn\ have been 
sufficiently resolved at radio frequencies to reveal morphological
details, namely SN~1986J (\cite{brs+91}) and
SN~1993J (\cite{bfr+94}; \cite{mar+95b}), both using VLBI (resolution 
$<1$ mas). While these supernovae are very different from \sn\ in their
luminosities, radio light curves and properties of their progenitor
stars (e.g.\ \cite{bk95b}), the brightness distribution of both objects 
deviates from that of a uniform spherical shell, at least to
some extent. Thus it is clear that \sn\ is not unique in having an
asymmetric radio morphology so early in its lifetime.  However, \sn\ {\em is}
unique in that we have a considerable amount of information
about the progenitor star and its environment which we can use to try
and explain the observed radio structures. Also, \sn\ appears be at a
later, or at least different, stage in its evolution: the other two
objects are still in the standard RSN phase, a phase which lasted just
days in the case of \sn.

Previous ATCA data published on \sn\ consists of 4.8 and
8.6~GHz flux densities up to day 2300, and a single super-resolved 9~GHz 
image. In this paper we present the entire ATCA
dataset, to date, on \sn. Flux density measurements have been carried out at
intervals of 1--2 months at 1.4, 2.4, 4.8 and 8.6~GHz from
day 918 (1989 August) up to day 3325 (1996 April).  These seven years
of observations cover the entire second phase of the radio evolution up
to this point. We have also obtained 9~GHz images of the remnant at four 
epochs from 1992 to 1995. With the increasing flux density of the remnant,
these images are becoming progressively more reliable and detailed.

Such observations provide a key insight into the interaction of the SN
shock with the CSM.  The flux density of the remnant as a function of time and
frequency is our best guide to conditions at the shock, such as the
details of the acceleration mechanism and the properties of the
material into which the shock is expanding. The radio images of the
remnant demonstrate the non-spherical nature of the system, and can be
interpreted in terms of the three-dimensional structure of the
surrounding medium and of the explosion itself. From these data the
true rate of expansion of the remnant can also be measured.  The
observations made in this project are discussed in
Section~\ref{Observations}. Results are presented in
Section~\ref{Results}, and are discussed in terms of the properties
of \sn\ and its surroundings in Section~\ref{Discussion}.

\section{Observations}
\label{Observations}

All observations were made with the Australia Telescope Compact Array
(\cite{fbw92}), an aperture synthesis telescope consisting of six 
22m-diameter antennae located on a 6km, east-west baseline near
Narrabri, New South Wales. Two types of observations have been made of
\sn\ using the ATCA: monitoring and imaging. All observations were
continuum observations, measuring two frequencies simultaneously each
with a bandwidth of 128~MHz.  All Stokes polarization parameters were 
recorded. Amplitudes were calibrated using the revised scale of 
Reynolds (1994)\nocite{rey94}, assuming a
flux density for PKS~1934--638 of 15.0, 11.6, 5.8 and 2.8~Jy at
frequencies of 1.4, 2.4, 4.8 and 8.6~GHz, respectively
(1~Jy $= 10^{-26}$ W m$^{-2}$ Hz$^{-1}$). Flux density measurements in
Papers I \& IV used an earlier scale;
measurements at 1.4, 2.4 and 4.8~GHz were on a scale
8--12\% higher, and those at 8.6~GHz 10\% lower, than values quoted
here. Phases were calibrated
using the source PKS~0530--727 and also, in conditions of poor phase
stability, PKS~0454--810 (see \cite{rjs+95}).

\subsection{Monitoring}

An ATCA monitoring observation of \sn\ normally gives a measurement
of the flux density of the source at four frequencies. The ATCA
can observe at two frequency bands simultaneously: earlier
observations were at either 1.4/2.4~GHz or 4.8/8.6~GHz, 
while more recent observations used frequency switching between
these two pairs. The integration time for an observation was typically 
6--12 h.

Until 1995 May, monitoring data were reduced using the AIPS package.
After this date, the Miriad package (\cite{stw95}) was used. To
check the consistency between these two, we reduced several observations
using both packages. The corresponding difference
in flux density was in all cases considerably less than the
uncertainty.

Data were edited, calibrated and imaged. Baselines shorter than 3~k$\lambda$ 
were often contaminated by emission from strong sources outside the 
primary beam (primarily 30~Doradus), and were discarded. Images were then
deconvolved using the CLEAN algorithm (\cite{cla80}). Before day 1500,
the flux density of \sn\ was equated to the peak brightness of the source.
After this date, the flux density was measured by fitting a Gaussian 
profile to the emission.

\subsection{Imaging}

Imaging observations were made three times a year from 1992 to 1995.
The three observations in a given year were always made with different
6km configurations to give increased radial coverage in the $u-v$
plane.  Each observation used two simultaneous frequencies, both between
8.6 and 9.1~GHz, giving a diffraction-limited synthesised beam of 
$\approx1''$. Integration times were long (12--16 h) to give increased
sensitivity. To avoid possible artifacts, the phase centre was placed
$10''$ south of \sn.  More information on these observations is
contained in Table~\ref{imaging}.

All imaging data were reduced using Miriad.  Observations corresponding
to earlier images (Paper III) were completely reprocessed.  Data
were carefully examined for bad points, and observations during which
the atmospheric phase stability was poor were rejected.  The data were
then calibrated and the three sets of observations for each year
combined.  Images centred on the remnant were produced using
multi-frequency synthesis (\cite{sw94}), 
using a cell size of $0''.08$ in each
coordinate and an image size of $512 \times 512$ pixels. 
As a compromise between natural and uniform weighting, Briggs' robustness
parameter was used with a value of 0 (\cite{bri95}).  This suppressed
stripes in the dirty map caused by the small cell size.  Deconvolution was
carried out using a maximum entropy (MAXEN) algorithm (\cite{gd78})
with 1000 iterations, using a square window of side $\approx 4''$
centred on the remnant.  The MAXEN model was then convolved with two
different restoring beams: the first was the diffraction-limited beam,
the second was a super-resolved circular beam with FWHM $0''.4$ as
used in Paper III.  This model was then
combined with the unscaled residuals, and then regridded at a pixel
scale of $0''.01$ in each coordinate to produce the final image.

\section{Results}
\label{Results}
\subsection{Monitoring}
\label{Monitoring}

Monitoring data for \sn\ at all four frequencies are shown in
Figure~\ref{light_curves}.  Two uncertainties are associated with
each measurement: a random error, representative of the noise
in the image and the uncertainty in the fitting process, and
a systematic error of order 5--10\%, 
estimated by examining the scatter in the flux density of a nearby ($\sim 4'$) 
unresolved source which we designate 0536--6919.
\footnote{This source is variable. However, the estimate of the
systematic scatter in its light curve
was made over a period of $\sim$400 d during which its
flux density appeared to remain constant.}
These errors are combined in quadrature to give the error
bars shown in Figure~\ref{light_curves}.

The flux density at all frequencies has continued to increase
essentially monotonically over the entire period of observations, 
and shows no sign of levelling off. A linear
fit over the last two years of data gives a rate of increase 
of 33.5$\pm$3.1, 23.0$\pm$3.0, 10.9$\pm$1.4 and
4.2$\pm$1.5  $\mu$Jy day$^{-1}$ at 1.4, 2.4, 4.8 and 8.6~GHz, respectively.
Short-time-scale variations are evident in the data --- the significance
of these is discussed in Section~\ref{short_term}.

Spectra for \sn\ at five epochs are shown in Figure~\ref{spectra}.
These observations are consistent with a constant spectral index 
$\alpha = -0.95\pm0.04$ over more than four years. 

\subsection{Imaging}

\subsubsection{Total intensity}
\label{imaging_section}

Two sets of total-intensity images are shown in Figures~\ref{1992_sn}
through~\ref{1995_sn}:  first the diffraction-limited images, then
the super-resolved images obtained by using a circular,
0$''$.4 FWHM restoring beam.  Only the inner $\approx 4''$ of each
40$''$ image is shown, and no emission
above the noise was seen outside this central region.  More information
on each image is given in Table~\ref{imaging}.
Figure~\ref{1995-1993_sn} shows an \sn\ difference image,
produced by subtracting the super-resolved image made in the
1993 epoch from that made in 1995.

In the diffraction-limited images of the supernova, the source is
clearly resolved. There is a distinct asymmetry associated with the
emission, in that the peak is displaced east of centre. While the
object increases in brightness with time, it does not appear to be
getting noticeably larger --- this is quantified in
Section~\ref{expansion}.  

The super-resolved images show that, rather than being centrally
peaked, the emission is restricted to the rim of an approximately
circular shell. The 1992 image is (unsurprisingly) similar to that
produced in Paper III:  two lobes are apparent on the east and west
limbs of the remnant.  In the 1993 and later images, the remnant
continues to brighten, and takes on a more circular shape.  For the
1995 epoch, the minimum of emission at the centre of the remnant is
330~$\mu$Jy beam$^{-1}$, well above the noise, and this can be considered an
upper limit on the flux density of a central point source.  
Using the CLEAN algorithm produces qualitatively similar
images, although it is worth noting that CLEAN is {\em not} recommended
for super-resolution (\cite{bri94}).

The difference image in Figure~\ref{1995-1993_sn} shows the two lobes or
``hotspots'' increasing in their intensity, the eastern hotspot clearly
increasing in brightness more rapidly than the western one. In 1992 the
ratio of peak brightness between the east and west lobes was 1.2, but
by 1995 it was 1.8. Approximately 50\% of the total flux density from the
remnant is contained in these regions.  In all images, the two hotspots
are at consistent position angles of $\sim100^{\circ}$ and $\sim260^{\circ}$.
The range in position angle of the diffraction limited beams in
Table~\ref{imaging} is large, and indeed the beams for the 1993 and 1994
epochs are nearly circular. This is good evidence that the consistent
orientation of the hotspots observed in the succession of super-resolved
images is not an artifact driven by the $u-v$ coverage of the observations.
Further tests on the reliability of such position angles under 
super-resolution have been carried out by Briggs (1995)\nocite{bri95}.

In the 1992 image, there appears to be an extension of emission towards
the north of the remnant.  This  vanishes in later images and is likely
to be an artifact introduced by super-resolution under conditions of low
signal-to-noise ratio (cf. \cite{bri94}).  In the 1994 and 1995 images, the
eastern hotspot seems to extend towards the south, and the difference
image in Figure~\ref{1995-1993_sn} indicates that this may be
a weaker, separate hotspot forming on the southern rim of the remnant.

Super-resolution is, in effect, an extrapolation of the {\em u-v} data
out to projected interferometer spacings beyond those measured by the
telescope. However, the long integration times, extensive {\em u-v}
coverage obtained and the simplicity of the structure in the resultant
images justify a moderate increase in resolution over that dictated by
diffraction theory (\cite{nn86}). The consistency of the results
further adds to the credibility of the method:  in 12 (three for
each year) independent datasets, each with different $u-v$ coverage and
with differing noise levels, the same features are reproduced in each
of Figures~\ref{1992_sn} through~\ref{1995_sn} --- a roughly circular
shell, with bright lobes to the east and west. 
For comparison, corresponding images of 0536--6919 are shown in
Figure~\ref{1995_src1}.  These show that for a simple unresolved
source, super-resolution behaves as expected, producing a Gaussian
region of emission with a reduced half-width and with an unchanged
total flux density. A more in-depth
study on the reliability of super-resolved images is described 
by Briggs (1994, 1995)\nocite{bri95}. 

Figure~\ref{hst_overlay} shows a comparison between the 
super-resolved radio image from 1995 and a WFPC2 HST image of the central
source and the surrounding inner optical ring (\cite{bkh+95}).
The images were aligned by first establishing an accurate
position for the central source
in the Hipparcos optical reference frame, and then
linking this with the radio reference frame as determined by
VLBI (\cite{rjs+95}). The residual uncertainty between the
VLBI reference frame and the ATCA image is $\sim 0''.05$.
The position angles of the two opposed brightness enhancements
in the radio image are similar to that of the major
axis of the optical ring (PA $\approx$ 81$^{\circ}$).
Although the minimum in the radio emission aligns closely with the
central source, this is not so in earlier images, and 
is not considered significant.

\subsubsection{Polarization}

Images were formed in the Stokes parameters Q, U and V, deconvolved
using the CLEAN algorithm and restored with a diffraction-limited
beam.  A linear polarization, L, image, was formed by combining Q and
U.  In the 1992, 1993 and 1994 data, no linear polarized emission was
detected from the radio remnant. In the 1995 data, there is a marginal
(2$\sigma$) detection from a region in the southwest of the remnant,
corresponding  to a total fractional polarization of $\sim$0.01\%. No
circularly polarized emission was detected at any epoch.

\subsubsection{Expansion of the remnant}
\label{expansion}

Using the method described in Paper II,
we obtain an estimate for the size of the radio remnant at each epoch
by fitting a two-dimensional projection of a thin, spherical shell 
to 9~GHz {\em u-v} data. We have used all the data shown in
Figure~2 of Paper III,
as well as that in Table~\ref{imaging}.
The results of the model fitting are shown in a plot of radius versus time
in Figure~\ref{radius_plot}. 
For the most recent (1995 November) data, a radius of
$r = 0''.685\pm0''.005$ is obtained,
84\% of the radius of the optical ring (\cite{bkh+95}).

A power-law fit to the data of the form $r/r_0 \propto (t/t_0)^m$
(where $t$ is the time since the SN explosion and $t_0 = 1800$~d)
gives $r_0 = 0''.641 \pm 0.006$ and $m = 0.12 \pm 0.02$.
At day 3200, this corresponds to an expansion rate of $9 \pm 2$
mas y$^{-1}$, which at a distance of 50~kpc translates to
2200$\pm$400 \km/s.  The value obtained for $m$ is slightly less, but
consistent with, the value $m = 0.17\pm 0.08$ obtained using data up to
day 2100 (Paper III).

A power-law fit implies that the shock is currently decelerating.
However, the data are equally well fitted by a
straight line.  This line does not pass through the origin, indicative
of a rapid deceleration in the past but minimal deceleration in the
present.  Such a fit gives an expansion rate of $12 \pm 2$ mas
y$^{-1}$ (2800$\pm$400 \km/s).  This is the value used in subsequent
discussion.

Modelling directly in the {\em u-v} plane is a
robust and consistent way of quantifying the relative size of the 
remnant as a function of time, and the results are not complicated
by issues of deconvolution and resolution. However, it is clear
that as time progresses, the approximation of a thin spherical
shell becomes less and less justified, and measurements of the
angular size may be skewed by the changing appearance of
the remnant. Therefore, to confirm these results, the size of the
remnant was also measured directly in the image plane.

For each super-resolved image of \sn, the radius of the remnant was measured 
along lines joining the position of the central optical source (\cite{rjs+95})
and the northern maximum, southern maximum, eastern saddle and western 
saddle, out to the point where the intensity
falls to 50\% of its value at the respective maximum or saddle. 
These measurements show no measurable expansion in any direction, with an
upper limit of $\sim$~7000 \km/s. Although this method is somewhat crude, 
it is consistent with the result obtained from the $u-v$ fitting above.

We find that the eastern side of the remnant is
$\approx 130 \pm 70$ mas further from the central source (in projection) 
than the other components. Reynolds et al. (1995)\nocite{rjs+95} 
noted a similar displacement by comparing the centre of symmetry of the 
1992 epoch image to the position of the central source.

\section{Discussion}
\label{Discussion}
\subsection{Flux density}
\label{monitoring}

\subsubsection{Long-term behaviour}
\label{long_term}

The second turn-on in radio emission from \sn\ (Paper I) has generally
been understood as being caused by the shock encountering a sudden
increase in the density of the CSM, presumed to be the reverse shock of
the BSG wind (\cite{che92b}; \cite{bbp94}). Blondin \& Lundqvist 
(1993)\nocite{bl93} point out the
possibility that no reverse shock exists, so that radio emission must
somehow be generated by density enhancements of unknown origin in the
free wind. However, the good fit to the radio light curve produced by
both simple (Paper IV) and detailed (DBK)
models, and the good
evidence presented in Sections \ref{expansion} and
\ref{expansion_discussion} for deceleration, are both strong arguments
in favour of the shock having encountered a region of increased density.

Models of such an encounter with the shocked BSG wind predict that the
flux density should level off or begin to decrease (\cite{che92b}; DBK). 
Yet it is clear from
Figure~\ref{light_curves} that the flux density continues to rise.  The
ellipsoidal geometry of the BSG wind cavity may be a contributing factor, in 
that the distance to the density jump is greatest along the symmetry
axis of the progenitor system (\cite{bl93}). The radio flare produced by
the encounter of the shock with dense material is consequently
spread out over a greater period of time before decaying.  However, the
morphology of the remnant suggests that emission along this axis does
not contribute significantly to the total flux (see further discussion below 
in Section~\ref{csm}).  Another possibility is that the presence of
toroidal magnetic fields within the BSG wind-bubble serves to boost the
radio emission (\cite{cl94}). This model can probably be ruled out though
by the low level of linear polarization observed, which argues against
the presence of magnetic fields with such a well-defined geometry
(Section~\ref{poln} below).
A recently proposed alternative is that the shock has encountered an
ionized region of the RSG wind located inside the optical ring
(CD95).  This \HII\ region is calculated to be $\approx400$ times
more dense than the stagnation zone of the BSG wind. The large
consequent deceleration experienced by the shock, combined with
the increased particle injection,
could sustain the radio flux density as is observed. 

We do not attempt to model the radio light curves, other than to note
that the simple model of Paper IV for flux densities up to day 2300 still
fits the data to a reasonable extent over the increased time period.
The morphology of the remnant is becoming increasingly asymmetric with 
time --- not only are there two strong hotspots, but they are brightening at
different rates, and new features may be forming.  Thus it is clear
that the assumption made in earlier analyses of a spherical emitting
region (DBK; Paper IV) is no longer valid, and that more complex
two- or three-dimensional modelling is now required.  Such a study is
beyond the scope of this paper.

\subsubsection{Short-term behaviour} 
\label{short_term}

In the ATCA data presented in Figure~\ref{light_curves}, small-scale
fluctuations in \sn's flux density are apparent at all four frequencies
over timescales of $\sim100$~d (comparable to the spacing between
observations).  This is in contrast with MOST data, for which Ball et
al. (1995a)\nocite{bcct95} conclude that there have been no variations
of amplitude $>$2\% on timescales between 30 and a few hundred days.
In Paper I, 4.8~GHz variations were reported at the level of $\sim$20\%
over a two-day period, six months after the radio emission reappeared.
However, since then the sampling interval has been far coarser, and we
are unable to determine if such short-period variability is still
occurring.

The nature of the ATCA means that flux densities are not as accurate as
those measured by the MOST. While the MOST has complete coverage in the
$u-v$ plane over a full 12-h synthesis and obtains all spatial
frequencies from $43\lambda$ to $4415 \lambda$, the ATCA has just
15 baselines spread over 6000 m. In addition, frequency switching
and phase calibration means that only about a third of the time can be
spent integrating on \sn\ at a given frequency.
These effects produce large sidelobes in
the resultant images. Figure~\ref{light_curves} suggests that 
flux density variations become more pronounced
at higher frequencies. However, this is likely the result of atmospheric 
phase variations which, at 4.8 and 8.6~GHz, can cause the signal to
decorrelate. There is the suggestion of some
correlations between frequencies which were observed simultaneously
(1.4/2.4~GHz or 4.8/8.6~GHz), but the factors discussed above
will affect the data at both such frequencies similarly.  We conclude that the
small-scale changes in flux density seen in ATCA data are probably not
significant --- the light curve obtained by MOST is far more reliable.

\subsubsection{Spectral index}

Observations in the first year after the radio emission re-appeared 
showed significant variations in the spectral index (Paper I).
Correcting for the ATCA revised flux density scale (\cite{rey94}),
the spectrum flattened from $\alpha < -1.6$ ($S \propto \nu^{\alpha}$)
on day 1243 to $\alpha = -0.91\pm0.05$ on day 1517, before possibly 
steepening to $\alpha = -0.99\pm0.04$ on day 1595.

The data in Figure~\ref{spectra} show that within the uncertainties,
the spectrum did not vary from day 1787 (1992 January) to day 3325
(1996 April). The average value over this period is $\alpha = -0.95\pm0.04$, 
consistent with the value at day 1595 (1991 July).

In the model of DBK, the shock compression ratio, $\rho = 1 -
3/2\alpha$, begins at its maximum value of 4 when the explosion occurs,
and consequently decreases because the shock is modified by the
acceleration of cosmic rays. The compression ratio then lies on a
plateau ($\rho \approx 2.7$) for an extended period before then rapidly
decreasing again.  In DBK's standard model, the plateau phase lasts
until approximately day 2000, and by day $\sim$3000, $\rho < 2$
($\alpha$ steeper than --1.5).  However, observations imply that on day
3325, $\rho = 2.58\pm0.11$.  For comparison, the spectrum on day 1517
corresponds to $\rho = 2.65\pm0.14$.

So, while we indeed observe a plateau, our results indicate that this
phase is considerably more extended than predicted, and that
accelerated protons are yet to smooth out the shock completely.
In DBK's model, the behaviour of $\rho$ as a
function of time depends on the distance from the progenitor star of
the density jump responsible for the radio emission. DBK's Figure 7
shows that if this jump is moved further from the star, the plateau
region does become more extended. The model of DBK assumes that the
density jump encountered is the termination shock of the BSG wind.
However, if, as discussed in Section~\ref{long_term}, 
the cause of the radio emission is an
ionized region of the RSG wind, the greater distance of this component
from the progenitor can, in the context of this model, 
explain the constant spectral index.
CD95 roughly estimate that the contact discontinuity with the
RSG wind is 20\% further from the star than the termination shock of
the BSG wind.

The increasingly asymmetric morphology of the remnant implies that a
direct interpretation of the spectral index in terms of a single shock
compression ratio may be overly simplistic. For example, because
structural information is only available at 9~GHz, we are unable to
determine if different regions in the remnant have different spectra,
as is observed in SN~1993J (\cite{mar+95a}). If this is the case, an
explanation for the observed constancy of the spectral index may require 
further insight.  

\subsection{Expansion of the remnant}
\label{expansion_discussion}

Between the SN event in 1987 February and the second radio
turn-on in mid-1990, 
a mean expansion rate can be calculated 
for the remnant of $\sim$35\,000 \km/s. This is consistent 
with free expansion since the explosion (\cite{hd87}).
However, by 1993 January, the remnant appeared to be
expanding at only 4800$\pm$2300 \km/s (Paper III). 
While this seemed to indicate considerable deceleration, there was 
only a 20-month period over which measurements of the remnant's size 
had been made, and the low signal-to-noise ratio meant that the errors in
these measurements were large. Also, an image was available at only 
one epoch, so that the possibility that the remnant was rapidly 
changing shape could not be ruled out.

However, over a period three times longer than previously possible,
a low expansion velocity of 2800$\pm$400 \km/s is still obtained.
The succession of images demonstrates that the shell 
does not appear to be changing in its gross form --- we therefore 
conclude that the explanation in Paper III for the observed deceleration
is incorrect, and that the expansion speed obtained is realistic.

One explanation for the extremely low expansion velocity is
that the emission observed is not from the shock at all, but from
stationary or slowly moving clumps of material in the BSG wind. In this
model, the shock is now considerably further from the star than
the remnant we observe.
However, the emission continues to increase at all radio frequencies,
and the spectral index is not steepening. Both these observations are
difficult to explain if electrons are not continuing to be accelerated 
in the emitting regions.  
We therefore favour the alternative, that
emission from the radio shell traces the expanding shock front.

The fastest ejecta contain only a small fraction of the total mass. For
example, in the model of Luo, McCray \& Slavin (1994)\nocite{lms94}, 
the ejecta have a density profile $n \propto v^{-9}$ for
$v >$~4\,000~\km/s. Thus it might be argued that while the
fastest ejecta are responsible for the radio turn-on at day 1200 (and
have since been decelerated), the main body of slower moving material 
has now caught up with this material, and that it is the
relatively unimpeded expansion of these ejecta which we are now
observing.  

In a simple calculation, $v$, the mean
velocity from day 0 to day $t$ of ejecta currently at radius $r$, is
given by
\begin{equation}
\label{min_velocity}
v = 18.6 \times 10^3~{\rm km~s}^{-1} \left( \frac{r}{0''.687} \right) 
\left( \frac{t}{3200~{\rm days}} \right)^{-1}
\end{equation}
where $r=0''.687$ is the angular size of the remnant at $t = 3200$ d,
using the linear fit in Figure~\ref{radius_plot}. Since in reality 
ejecta may have decelerated in reaching a radius $r$, $v$ is a {\em lower
limit} on the initial velocity of the ejecta currently at this distance. 

From Equation~(\ref{min_velocity}), we see that while the initial
radio turn-on ($r \approx 0''.5$, $t \approx 1200$ d) requires
ejecta of velocity $\sim$35\,000 \km/s, emission from a thick shell at
day 3200 still requires ejecta for which
$v~=$~10\,000~--~20\,000~\km/s.  The consequence of this is that most
of the mass ejected by the supernova cannot yet have reached the radio
emitting regions --- the remnant is the result of rapidly moving
material which was significantly decelerated at some time in the past,
presumably around the time that radio emission was re-detected.
Figure~\ref{radius_plot} is consistent with minimal deceleration from
day 1500 to day 3500, but we cannot rule out the possibility that
deceleration continues to take place.

This apparent deceleration would appear to be the best observational
evidence yet that the shock has encountered a marked increase in
density. CD95\nocite{cd95} argue that this deceleration is further evidence 
for an encounter with the dense, ionized RSG wind. From their model,
we calculate a {\em mean} velocity for the shock while it is in this region of
6100$\pm$1200 \km/s. If the remnant continues to decelerate, 
a still denser component may be required to explain the observations.

When the fastest ejecta reach the circumstellar ring, the system is
expected to become a strong source of optical and UV line emission, and
the X--ray and radio luminosity should show a rapid increase
(\cite{lm91b}; Luo et al. 1994 \nocite{lms94}).  
This has clearly not yet occurred, and
limits even the fastest ejecta to a mean velocity of $\approx$22\,000
\km/s from day zero to day 3200. Assuming that the remnant continues to
expand at its current rate, our results indicate that the shock will
reach the ring in the year 2006$\pm$3, which agrees with the prediction
of 2003 from X--ray observations (\cite{hat96}). This is also consistent with
hydrodynamic models (e.g.\ \cite{ssn93}; Luo et al. 1994), the most recent
estimate being 2005$\pm$3 (CD95). However, further deceleration of the
ejecta could delay the encounter considerably.

In Section~\ref{expansion}, we suggested that the eastern region
of the remnant may be further from the centre of the explosion than
other regions.  If this is the case, then this region either had the highest 
initial velocity, or has been decelerated the least. That it is also the
brightest region favours the former.  This could represent
anisotropy in the explosion, as discussed below in
Section~\ref{explosion}.

\subsection{Polarization}
\label{poln}

The radio emission from the remnant is believed to be synchrotron
emission (\cite{bk92b}; Paper IV),
and so is expected to be associated with
high degrees of linear polarization. However, the remnant is
comparable in size to the synthesised beam of the telescope, and so
there may exist variations in position angle of the electric field
vectors over spatial scales smaller than this beam.  In addition, the
projected direction of the magnetic field could also be varying along
the line of sight.  Both these effects can smear out any linear
polarization and give a low value for L. From the low level of
polarization observed, we can conclude that the magnetic field is
disordered on a scale of $\sim0.5''$ (0.1 pc), and certainly is not
well organised within each hotspot.

\subsection{Morphology of the radio emission}

It is well established that the triple-ring optical nebula surrounding
the supernova has cylindrical symmetry, with the axis
inclined at $\approx 43^\circ$ to the line of sight and 
at a position angle of $\approx -9 ^\circ$ on the plane
of the sky (\cite{jab+91}; \cite{bkh+95}; \cite{plck95}).
The supernova event itself also appears to have
had a similar symmetry, the envelope being extended along the axis
of the cylinder (\cite{kkn+89}). It is generally accepted
that the axisymmetric geometry observed coincides with the axis 
of rotation of the progenitor star, or possibly with the axis
of a progenitor binary system.

An optically thin cylindrical shell of radio emission inclined
at $\sim 45^{\circ}$ to the line of sight will 
show two brightness enhancements on the inclination axis.
Thus in Figure~\ref{hst_overlay}, the alignment of the two opposed radio
hotspots with the major axis of the optical ring argues that, to first
order at least, the radio emission has the same  
symmetry as the optical nebula.  However, the enhanced brightness from
the eastern limb of the radio remnant breaks this symmetry, and
indicates that the true geometry is still more complex.  Models of the
flux density cannot distinguish between ring and spherical geometries
for the radio emission (Ball et al. 1995b\nocite{bcs95}).  
However, the non-zero level of
emission in the centre of the radio shell argues that, unlike the inner
ring, the remnant is quasi-spherical and extended in three dimensions.

DBK attribute the anisotropies observed in the 1992 image to
minor variations in conditions at the shock, and assume that the system
is essentially spherically symmetric.  However, the persistence of these
anisotropies, and the increase of the asymmetry with time, argue that
the non-spherical nature of the morphology is significant.  Bartel et
al. (1991)\nocite{brs+91} suggest three explanations for the
asymmetries observed in the radio remnant of SN~1986J:  a central
pulsar could be affecting the morphology in some way, the explosion may
have been asymmetric, or the CSM into which the shock is expanding is
not isotropic. Each of these could also influence the evolution of
SN~1987A, and we consider the three possibilities in turn.

\subsubsection{Effect of a central pulsar}
Neither pulsed emission (e.g. \cite{mp96}) nor a central compact
source (Paper I)
has yet been detected in the remnant of \sn. Although the detection of
neutrinos from the initial explosion suggests that a neutron
star was formed, it may have undergone subsequent collapse to
form a black hole (\cite{che92}). If this is not the case, effects such 
as luminosity, beaming, scattering and absorption may still prevent
the detection of pulsed emission (\cite{man88}).

While the radio emission observed is clearly not ``plerionic'',
it has been proposed that the formation of two opposed regions
of enhanced emission in shell supernova remnants (SNRs) 
can be explained by jets or cones of
emission from a central neutron star (\cite{man87}; \cite{wwps96}). 
However, the lack of radio emission
for several years between the SN event and the second turn-on,
the steep spectral index,
and the increasing asymmetry between the eastern and western
hotspots cannot be explained by this model in its simplest form, 
and we consider it an unlikely explanation for the radio emission.  
It is worth noting that a model involving a pulsar with precessing
jets has been proposed to explain the optical triple-ring nebula
(\cite{bk95a}).

\subsubsection{An asymmetric explosion}
\label{explosion}
There is reasonable evidence that supernova ejecta are not
distributed isotropically.
Models of SNe involving rotating progenitor stars produce 
excess ejecta in the equatorial plane (\cite{bis71}; \cite{bw83}), and there
is observational evidence to suggest that this can influence
the morphology of an evolved SNR (\cite{tcb82}; \cite{kc87}). 

For a spherically symmetric ambient medium with no large-scale magnetic
fields, the radio emission from the remnant will peak where the shock
is strongest.  Indeed, several authors have argued for a 
spherically symmetric BSG wind (e.g.\ \cite{ma95}) and the lack
of polarization detected in the radio emission indicates that the
magnetic field is disordered on small scales.  With such simplifying
assumptions, the alignment of the radio morphology with the presumed
equatorial plane of the progenitor star suggests that the ejecta were
predominantly toroidal. In addition, the shell is faintest in the north
and the south, as would be expected if ejecta were primarily generated
in the equatorial plane.

There are problems with this conjecture, however. First, the
suggestion that a third, southern, hotspot may be forming in the shell (as 
shown in Figure~\ref{1995-1993_sn}) is difficult to explain.  Second, the
assumption that the CSM which the shock is encountering is spherically
symmetric may well be a naive one, as discussed further in
Section~\ref{csm}.  Finally, although there is much evidence from
both optical polarimetry and asymmetry in the line emission from the
supernova to suggest that the envelope and ejecta were axisymmetric 
(\cite{cbm+88}; \cite{mcb+88}; \cite{kkn+89}; \cite{pkcg95}), 
most observations have
extensions being produced predominantly in the {\em polar} directions.
This is at odds with the equatorial dominance required here.  However, the 
asymmetries represented in optical observations may only be of very
low mass (e.g.\ \cite{uca95}) and may not be representative of the
overall shock front.  VLBI observations of SN~1993J have indeed shown
that early measurements of asymmetry in the SN event need not translate
to the morphology of the radio shell (\cite{mar+95b}).  Also, simulations
of axisymmetric ejecta show that the resulting shock front is almost
spherical (\cite{blc96}). Given these results, we cannot
completely rule out this hypothesis.

Rather than producing a toroidal ejection of material, SNe can also be
directional in their anisotropy, as evidenced by the kicks imparted to
neutron stars (\cite{kbm+96}; \cite{pbw96}). In
Section~\ref{expansion}, it was shown that the eastern limb may have
expanded slightly further than the western limb, which may be the
result of more rapidly moving material being ejected in this
direction.  Note that optical spectroscopy of
\sn\ also gives evidence for such an event (\cite{sta96}).

\subsubsection{Asymmetries in the CSM}
\label{csm}

The nature of the triple-ring nebula surrounding \sn\ is not completely
understood (\cite{bkh+95}; \cite{bk95a}; \cite{psg+96}).  However, it
is generally agreed that its existence is consistent with the mass-loss
history of the progenitor star Sk--69$^\circ$202.  Several authors
have proposed as an explanation the interaction of an equatorially
enhanced RSG wind with a spherically symmetric BSG wind (\cite{lm91b};
\cite{wm92}; \cite{ma95}; CD95\nocite{cd95}).  Other possibilities
include a spherically symmetric RSG interacting with an axisymmetric
BSG (\cite{blo94}), the effect of a binary companion (\cite{pfs91};
\cite{lok95}) and the presence of a proto-stellar molecular disk
(\cite{ml94}).  In any case, it is accepted that the surrounding CSM is
not spherically symmetric.

Chevalier \& Luo (1994)\nocite{cl94} have argued that the
onset of radio emission from \sn\ can be explained by toroidal magnetic
fields in the BSG wind bubble. Only a slight enhancement in the magnetic 
field of the equatorial regions above the
poles is required to produce the observed morphology (DBK). But as discussed
in Section~\ref{poln}, the lack of observed polarization makes this
explanation unlikely.

A plausible explanation for the hotspots observed in the radio emission
is that these parts of the shock have simply encountered denser regions
of the CSM than the fainter regions. This could be because the
transition to the shocked wind occurs closer to the star in the bright 
regions, or because the density of the shocked wind varies with polar
angle.  These two possibilities are not mutually exclusive, and we
consider the implications of each below.

If the termination between the free and shocked winds is not spherical,
but is extended along the symmetry axis (\cite{bl93}; CD95), the
equatorial regions of the SN shock encounter the denser, shocked wind
before the polar regions do. The polar regions can also be expected 
eventually to reach the shocked component of the wind, at which time the
remnant might begin to take on a more uniform brightness distribution.
The suggestion in the 1994 and 1995 super-resolved images
(Figures~\ref{1994_sn} and~\ref{1995_sn}) and in the difference image
(Figure~\ref{1995-1993_sn}) that the southern rim of the radio shell is
now beginning to brighten may indicate that this is already happening.

The fact that the eastern hotspot is brighter than its western
counterpart can be understood if this side of the expanding remnant
reached the density jump first.  This may correspond to the two-clump
model of DBK, where the SN shock encounters two clumps of material with
a $\sim$120 d delay between them. No third jump in the light curve is
seen which might correspond to the formation of a new hotspot to the
south (A.J. Turtle, private communication), although its contribution to the
total flux density would be low.  This difference between the eastern and
western sides of the remnant can be explained in terms of a break in
the cylindrical symmetry of the surrounding CSM, although this could be
at odds with the high degree of symmetry observed in
the optical emission (\cite{bkh+95}). Light travel times may also have 
an effect (Ball et al. 1995b\nocite{bcs95}), 
but a more appealing prospect is that it is a
result of the possible directional anisotropy in the initial explosion
discussed in Section~\ref{explosion}. If higher velocity ejecta
were produced on the eastern side of the remnant, this can also explain 
why the eastern hotspot continues to brighten more rapidly
than that in the west.

While the shocked wind is believed to have a constant density in the
radial direction (e.g.\ DBK), it has been argued that the density may
well vary with polar angle (\cite{blo94}), so that even after all parts
of the SN shock have reached the denser component of the wind, particle
injection may be greatest around the equator.  Such an axisymmetric density
distribution in the CSM should produce extensions in the north-south
direction (Paper III; Blondin et al. 1996\nocite{blc96}).  
Although such extensions are
clearly not observed, Blondin et al. estimate that
protrusions along the poles only become significant $\sim$10~yr
after the explosion, and so may not necessarily be expected to occur at
the current time.

It is worth noting that the amount of anisotropy in the CSM required to
produce the observed morphology is low. Around any given circumference
on the shell in the 1995 super-resolved image (Figure~\ref{1995_sn}),
the ratio of maximum to minimum brightness is no more than three.
Using Equation~(4.7) of Paper IV, this corresponds to a ratio of the ambient density 
between the equatorial and polar regions of $\sim$1.7.
Most models can easily account for such a difference, even if the shock
is yet to reach the densest regions (\cite{lm91b}; \cite{bl93}).

\subsubsection{Comparison to other RSNe and SNRs}

SN~1986J, one of the most luminous RSNe known,  was imaged in 1988 by
Bartel et al. (1991)\nocite{brs+91}. It was then of
comparable age to \sn, and also had a radio remnant quite similar in
morphology; it too had two opposed hotspots, one brighter than
the other, superimposed on a rough circular shell.  SN~1986J also has
several protrusions extending beyond the shell.  These appear to extend
from local minima in the rim, and might be representative of the
features discussed by Blondin et al. (1996)\nocite{blc96} which are
produced by the expanding SN shock in the less dense, polar, regions of
an axisymmetric progenitor wind. As noted above, such features may, with
time, become apparent in \sn.

Marcaide et al. (1995b)\nocite{mar+95a} present a 
time sequence of images of the
remnant of SN~1993J, taken between 6 and 18 months after
the explosion. Unlike \sn, clear expansion has been 
detected for this object --- an expansion index of
$m \approx 0.9$ ($r \propto t^m$)
has been measured right from day zero, indicating minimal
deceleration. At 8.3~GHz, the shell of SN~1993J is almost 
circular, and has relatively small 
(compared to \sn) brightness variations around
its circumference. Marcaide et al. (1995a)\nocite{mar+95b} attribute
these variations to small inhomogeneities in the CSM, possibly
produced by a binary companion to the progenitor. 

In SN~1993J, we have a remnant which has so far
undergone minimal deceleration, and which is also reasonably uniform
in its shape and brightness. If, from this, we argue that it is the same
physical process which causes both the deceleration of a shock front
and a non-uniform appearance for the radio remnant, then we are led to
the conclusion that it must be a non-uniform CSM which gives \sn\ its
appearance.

The connection between radio supernovae and evolved supernova remnants
is unclear (\cite{wsp+86}; \cite{wd90}), and there is the intriguing
possibility that having long ago passed through its RSN stage, \sn\ may
be, as suggested in Paper III, in the very early stages of an evolution
corresponding to that of a classical SNR.  Manchester
(1987)\nocite{man87} has argued that the passage of the SN shock
through the progenitor wind, as is occurring in the case of \sn, can
``imprint'' a remnant with a morphology which it retains through to an
evolved state.  For example, the small-scale symmetry of
SNR~G296.5+10.0 is indicative of the remnant's being shaped by events
early in its lifetime (\cite{ssmk92}).  However, it is clear that
the large-scale ISM (\cite{lprv82}; \cite{bs86}), as well as possibly
large-scale magnetic fields (\cite{rmk+88}; \cite{fr90}), may both have
a large effect on a remnant's morphology.

At low resolution, the bipolar, but asymmetric, appearance of older
supernova remnants such as SNR~G308.8--0.1 and SNR~G320.4--1.2
(\cite{wg96}) is suggestive of the morphology of \sn.  
Such a comparison is, at least for now, pure speculation, 
although continued observations of \sn\ over
the next decade and beyond may give hints as to what this object might
one day become.

\section{Conclusion}

In this paper, we have reported on all ATCA observations of \sn\
from 1989 up to mid-1996, with the following
main results:

\begin{enumerate}
\item The total flux density has increased monotonically since radio emission
was redetected in 1990.
\item The spectral index has remained constant since 1992 January 
at $\alpha = -0.95 \pm 0.04$.
\item Assuming that the observed radio emission traces the shock,
the remnant is the product of initially rapidly moving ($v =$ 10\,000
-- 35\,000 \km/s) ejecta, which have subsequently decelerated 
to $v \approx 3000$ \km/s. Assuming continued expansion at this
velocity, the shock will encounter the inner optical ring in the 
year 2006$\pm$3.
\item The eastern and western regions of the remnant align with the major
axis of the optical ring, and are brighter than 
the northern and southern regions.
\item The eastern limb is brighter than the western limb, and is
also increasing in brightness more rapidly.
\item Material on the eastern side of the remnant may have travelled
further from the explosion than the rest of the remnant.
\end{enumerate}

We propose the following basic theory. Points 1, 2 and 3 above are
consistent with the shock reaching a sudden increase in the density of
the CSM.  This is believed to be either the termination shock of the
BSG wind, or the contact discontinuity to the RSG wind. The increased
density and distance from the star of the latter region can perhaps
better explain the observations.

Point 4 is good evidence that the shocked wind is axisymmetric in its
shape and/or density distribution.  That deceleration of a remnant
seems coupled with distortion of its morphology (cf.\ SN~1993J) is also
a good indication that the CSM is probably the cause for the observed
structure.  We expect that as the shock further expands into this
axisymmetric CSM, the polar regions will elongate and eventually
brighten. If observations show these regions remaining faint and
becoming retarded in their expansion, an alternative explanation could
be a toroidal distribution of ejecta.

Points 5 and 6 can be explained by the shock on the eastern side of the
remnant reaching the density jump first. We propose that this
may indicate a directional anisotropy in the SN explosion.

Monitoring and imaging \sn\ with the ATCA will continue.
In addition, the forthcoming upgrade of the ATCA to
operation at frequencies up to 25~GHz 
will allow diffraction-limited images of \sn\ down to a
resolution of $\sim0''.4$. With such data, a better study of existing
and new features will be possible.  With resolved images at both 9 and
25~GHz, spectral index variations across the remnant will also be
detectable, which may show differences in the energetics of hotspots
and of the polar regions. For example, the brightest region of
SN~1993J also appears to have the steepest spectrum (\cite{mar+95a}).

It is not clear exactly what will happen to the radio emission when the shock
reaches the optical ring --- much will depend on the nature of the
unshocked RSG wind and its density distribution, an issue on which
current models are undecided. It is entirely possible that the
appearance of the remnant once it has passed into this region will be
quite different to what we see now.  In any case, \sn\ will likely
become a spectacular object in all wavebands, and we will continue to
observe it regularly in anticipation of this event.

\acknowledgements
We thank Lewis Ball, Dan Briggs, David Crawford, Bob Sault, Raylee Stathakis, 
Tony Turtle and Mark Walker for useful discussions and
advice, and Chris Burrows and John Krist for their HST image of the
optical nebula. We are also particularly grateful for the assistance,
support and patience of the staff of the Paul Wild Observatory during
the large number of observations for this project over the last seven
years. BMG acknowledges the support of an Australian Postgraduate
Award. 

Requests for data should be directed via email to
b.gaensler@physics.usyd.edu.au.

\newpage

\begin{table} 
\begin{tabular}{llccccc} \hline \hline
Epoch & Date of     & Day No. & Integration & Diffraction  & RMS Noise         & S  \\
      & Observation &   & Time        & Limited Beam & ($\mu$Jy/beam) & (mJy) \\  \hline
1992  & 21-Oct-1992 & 2068 & 17h & $1''.11 \times 0.''85$ & 44 & 5.4$\pm$0.1 \\
      & 04-Jan-1993 & 2142 & 15h & PA $= 34^{\circ}$ & & \\
      & 05-Jan-1993 & 2143 &  7h & & & \\ \hline
1993  & 24-Jun-1993 & 2314 & 13h & $0''.98 \times 0''.92$ & 34 & 6.8$\pm$0.1 \\
      & 01-Jul-1993 & 2321 & 13h & PA $= 12^{\circ}$ & & \\
      & 15-Oct-1993 & 2426 & 19h & & & \\ \hline
1994  & 16-Feb-1994 & 2550 & 13h & $0''.99 \times 0''.94$ & 30 & 7.3$\pm$0.1 \\
      & 27-Jun-1994 & 2683 & 26h$^*$ & PA $= -78^{\circ}$ & \\
      & 01-Jul-1994 & 2687 & 14h & & & \\ \hline
1995  & 24-Jul-1995 & 3074 & 11h & $1''.04 \times 0''.89$ & 33 & 10.8$\pm$0.1 \\
      & 29-Aug-1995 & 3111 & 11h & PA $= 6^{\circ}$ & & \\ 
      & 06-Nov-1995 & 3178 & 12h & & & \\ \hline
\end{tabular}
\\
\\
\footnotesize
$^*$spread over two days
\normalsize 
\caption{Imaging observations of \sn\ with ATCA. Position angles are those 
of the major axis of the beam, and are measured north through east.}
\label{imaging}
\end{table}

\bibliographystyle{apj1}
\bibliography{modrefs,psrrefs}

\newpage

\noindent
FIGURE CAPTIONS \\ \\
Figure 1 : Light curves for the second radio turn-on of \sn\ at
four different frequencies. Where error bars are not shown,
they are smaller than the plotted points.
 
\noindent
Figure 2 : The radio spectrum of \sn\ at five different epochs.
The day number corresponds to the number of days since the explosion, 
and all observations were made within three days of the day number
on each plot. Where available, 0.843~GHz MOST data (\protect\cite{bcct95}) 
have been included. The spectral index~$\alpha$ ($S \propto \nu^{\alpha}$) 
has been calculated using a linear least-squares fit. Errors are
no larger than the size of the plotted points.

\noindent
Figure 3 : Diffraction-limited (top) and super-resolved images (bottom) of 
\sn\ at 9~GHz, using data from epoch 1992 (see Table~\ref{imaging}). An ellipse
corresponding to the FWHM of the Gaussian restoring beam is shown at
bottom right of each image. For the diffraction-limited image, the
greyscale runs from --0.1 to 5.0 mJy beam$^{-1}$, and the contours run from
0.5 to 5.0 mJy beam$^{-1}$ at 0.5 mJy beam$^{-1}$ intervals. 
For the super-resolved
image, the greyscale is from --0.1 to 2.0 mJy beam$^{-1}$, with contours
from 0.2 to 2.0 mJy beam$^{-1}$ at 0.2 mJy beam$^{-1}$ intervals.

\noindent
Figure 4 : As in Figure~\ref{1992_sn}, for 1993 data.

\noindent
Figure 5 : As in Figure~\ref{1992_sn}, for 1994 data.

\noindent
Figure 6 : As in Figure~\ref{1992_sn}, for 1995 data.

\noindent
Figure 7 : A difference map for \sn\ between 1993 and 1995.
The greyscale corresponds to the super-resolved 
1995 epoch data for \sn\ (as in Figure~\ref{1995_sn}), and ranges from 
--0.1 to 2.0 mJy beam$^{-1}$. The contours correspond to a difference image 
between the 1995 and 1993 super-resolved data. Broken contour levels are
--0.2 and --0.1 mJy beam,$^{-1}$ while solid levels are 0.1 to 
0.9 mJy beam$^{-1}$ in increments of 0.1 mJy beam$^{-1}$.

\noindent
Figure 8 : Diffraction-limited and super-resolved images of 0536--6919 at
9~GHz, using data from epoch 1995. For the diffraction-limited image, the
greyscale runs from --0.1 to 6.0 mJy beam$^{-1}$, and the contours run from
0.5 to 6.0 mJy beam$^{-1}$ at 0.5 mJy beam$^{-1}$ intervals. 
For the super-resolved
image, the greyscale is from --0.1 to 5.0 mJy beam$^{-1}$, with contours
from 0.5 to 5.0 mJy beam$^{-1}$ at 0.5 mJy beam$^{-1}$ intervals.

\noindent
Figure 9 : A comparison of HST and ATCA images of \sn.
Contours show the 1995 epoch of the super-resolved ATCA 9~GHz data. Contours 
run from 0.2 to 2.0 mJy beam$^{-1}$ at 0.2 mJy beam$^{-1}$ increments. At 
bottom right is shown the $0''.4$ FWHM restoring beam. The greyscale is 
a WFPC2 HST image of \sn\ in H$\alpha$ + [N{\sc ii}] (\protect\cite{bkh+95}),
to which the reference frame 
of Reynolds et al. (1995)\protect\nocite{rjs+95} has been applied.
The point source at the lower left of the HST image is an unrelated
field star (star 3 of \protect\cite{ws90}).

\noindent
Figure 10 : The radius of the radio remnant as a function of time,
as determined by fitting a thin spherical shell of arbitrary
position, flux density and radius to the $u-v$ data for 9~GHz observations
in Paper III
and in Table~\ref{imaging}. The broken line is a power-law
fit to the data of the form $r \propto t^m$, with $m = 0.12$. The
solid line is a linear fit to the data, with slope
32~$\mu$as~day$^{-1}$ and with $r = 0''.642$ at $t = 1800$ days.

\newpage

\begin{figure}
\centerline{\psfig{file=fluxes.ps,height=20.0cm}}
\caption{}
\label{light_curves}
\end{figure}
 
\begin{figure}
\centerline{\psfig{file=spectra.ps,height=20.0cm}}
\caption{}
\label{spectra}
\end{figure}

\begin{figure}
\centerline{\psfig{file=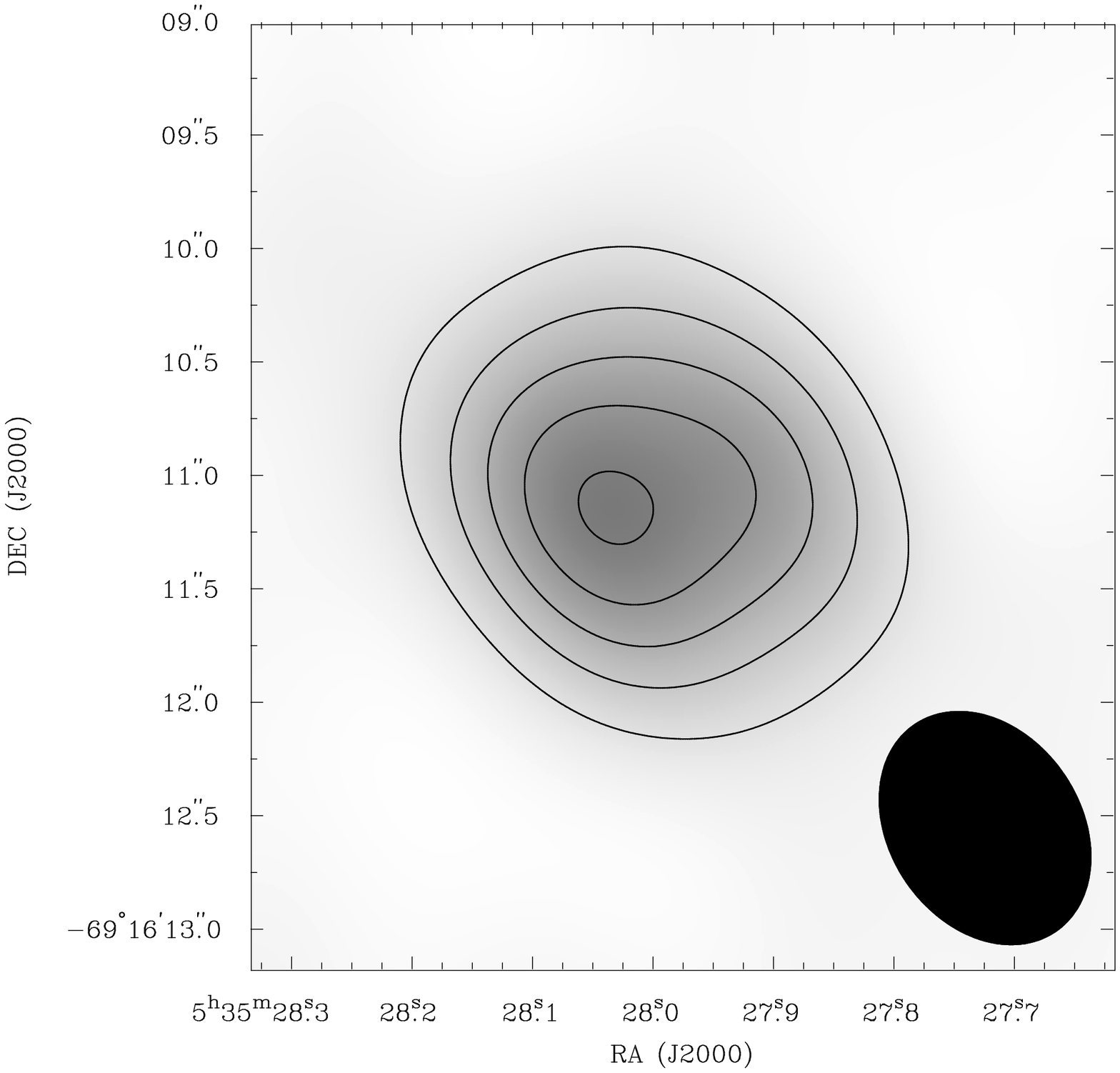,height=10.0cm,angle=270}}
\vspace{2cm}
\centerline{\psfig{file=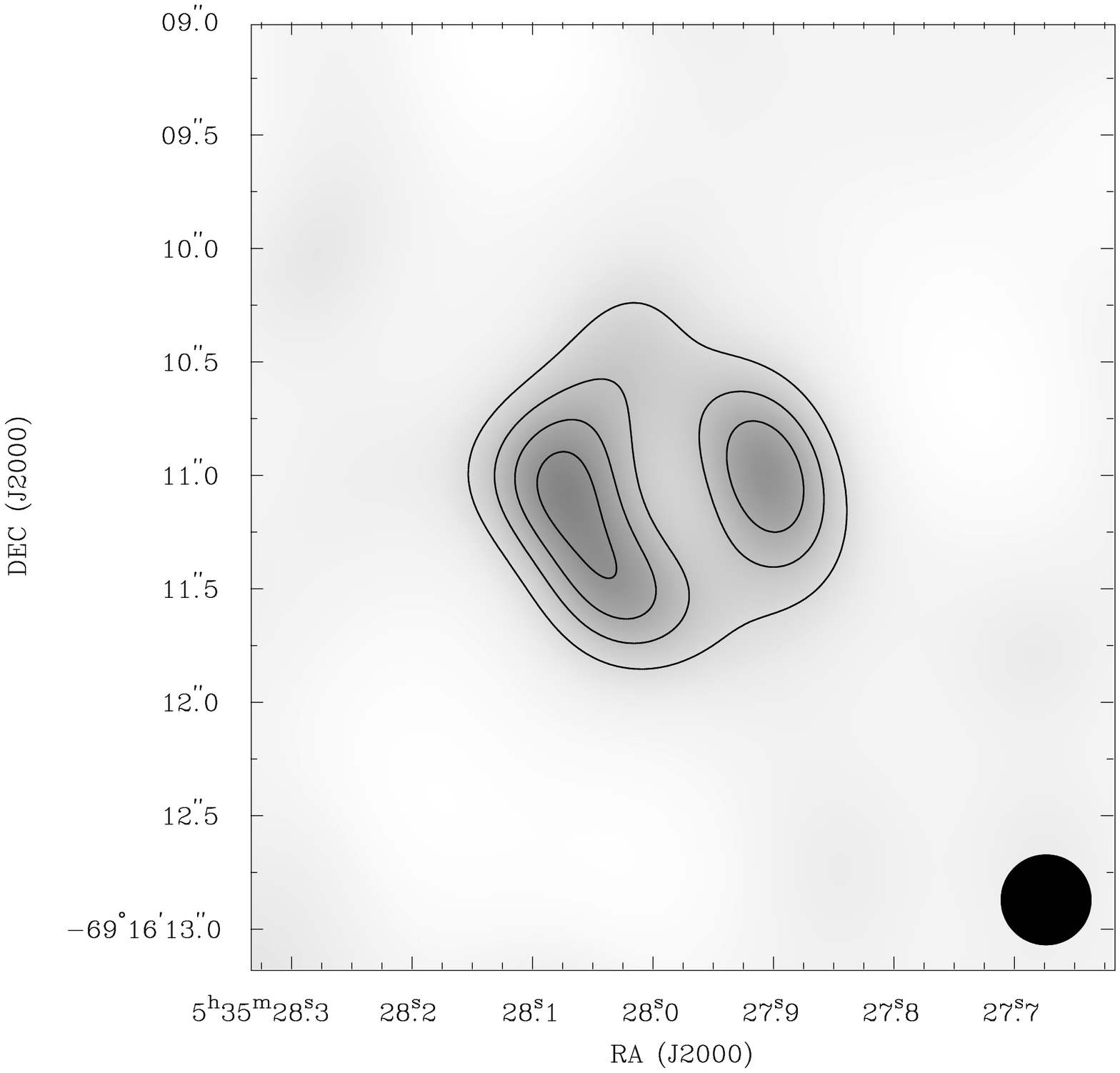,height=10.0cm,angle=270}}
\caption{}
\label{1992_sn}
\end{figure}

\begin{figure}
\centerline{\psfig{file=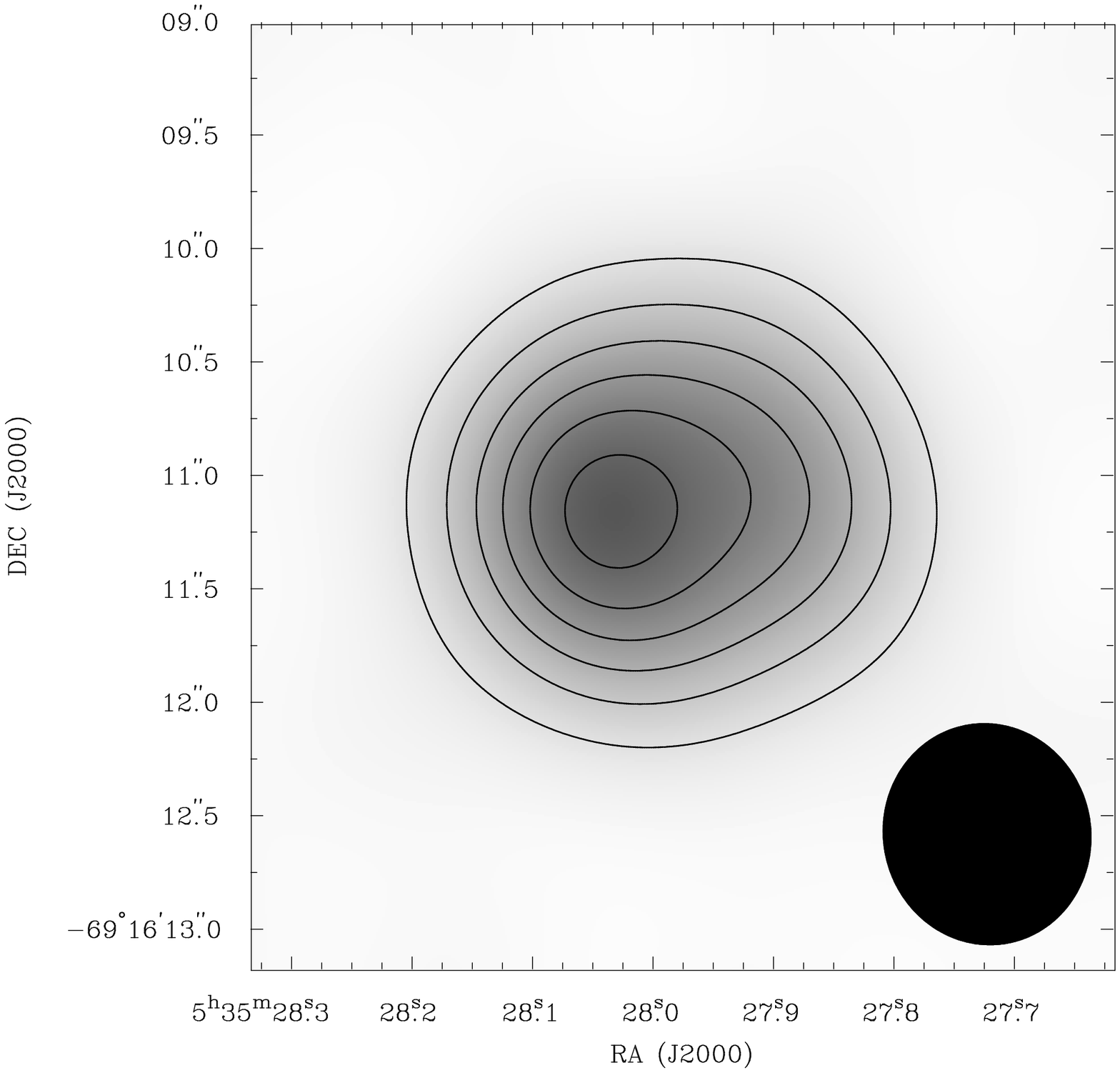,height=10.0cm,angle=270}}
\vspace{2cm}
\centerline{\psfig{file=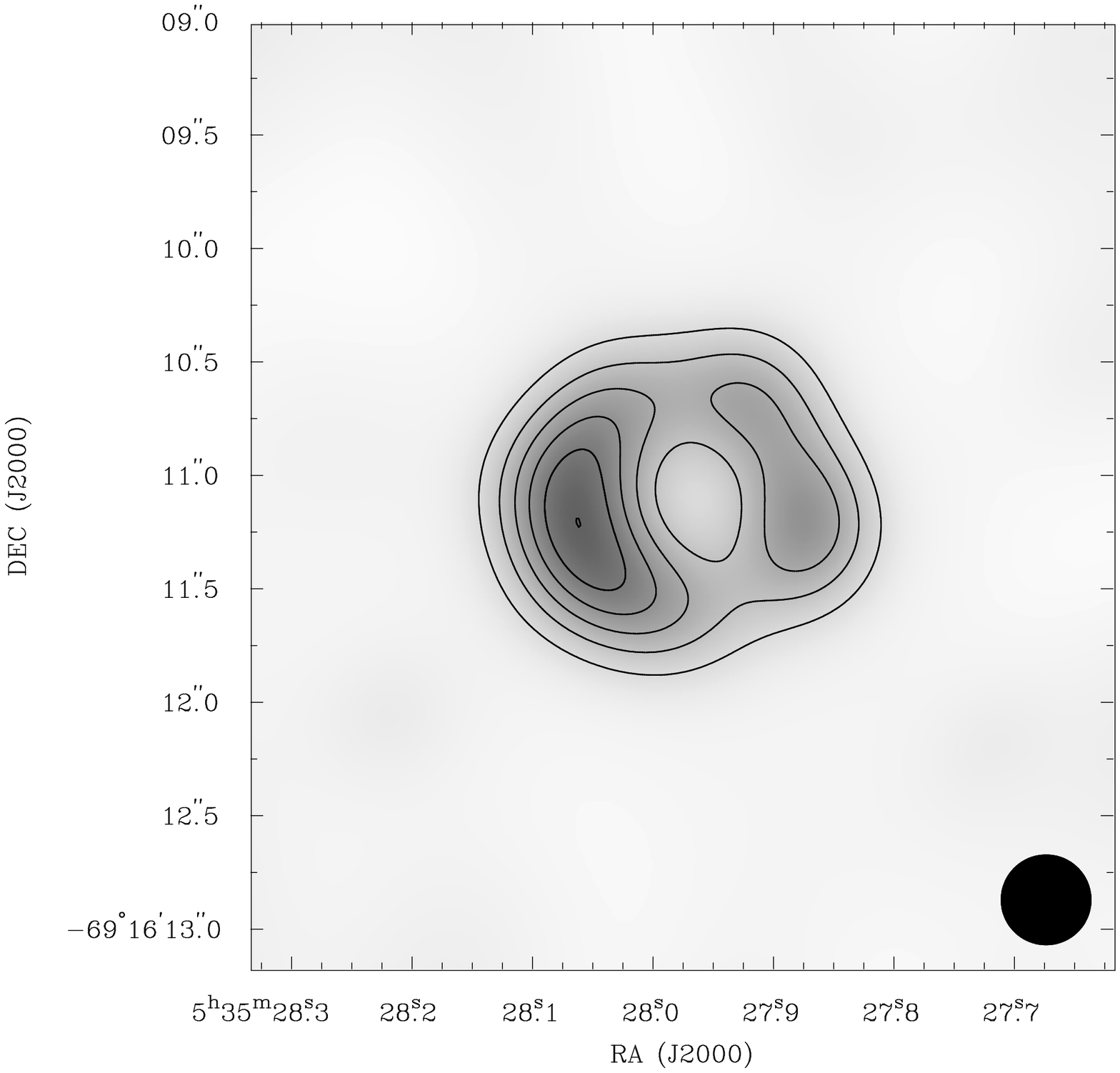,height=10.0cm,angle=270}}
\caption{}
\label{1993_sn}
\end{figure}

\begin{figure}
\centerline{\psfig{file=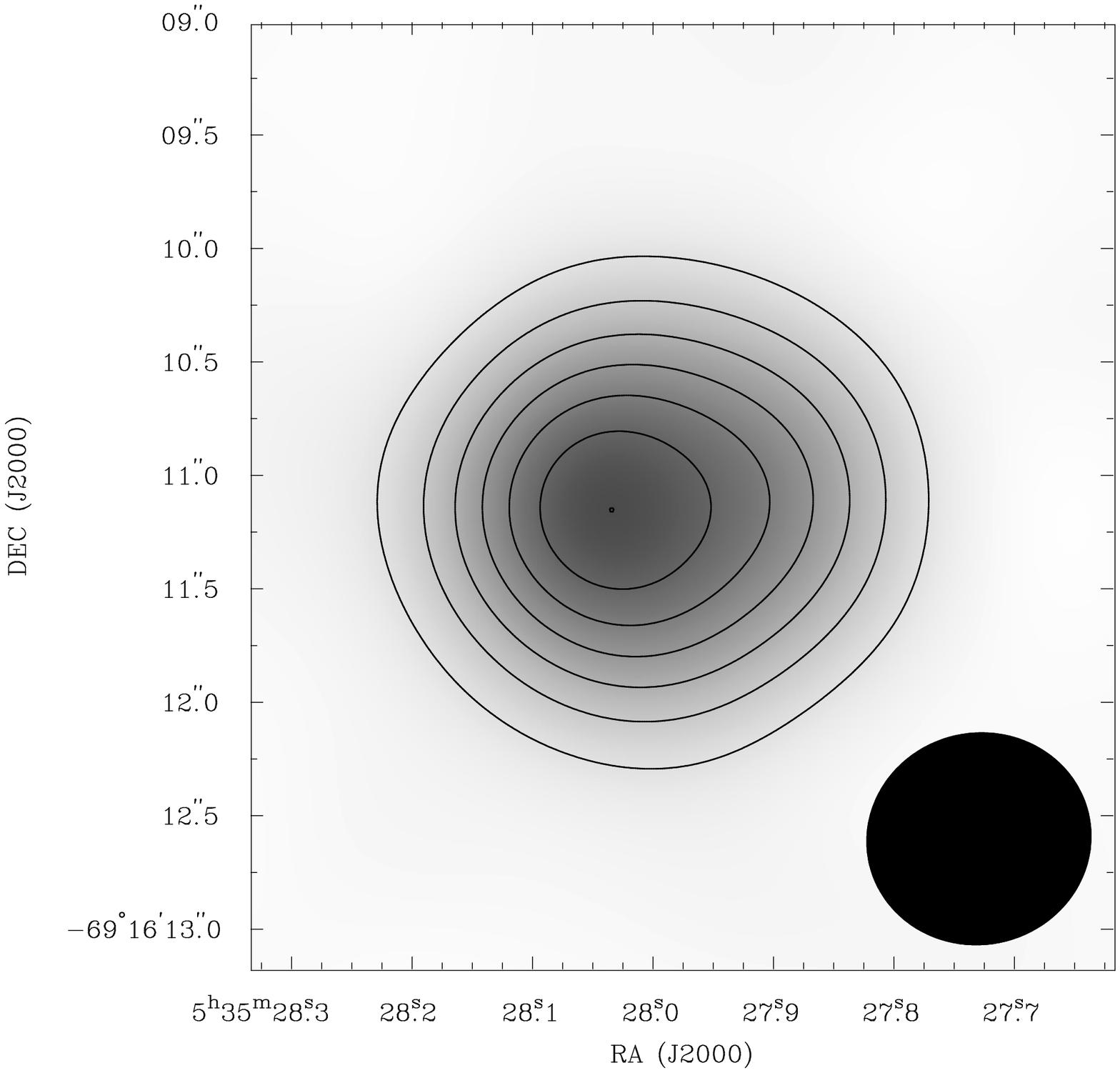,height=10.0cm,angle=270}}
\vspace{2cm}
\centerline{\psfig{file=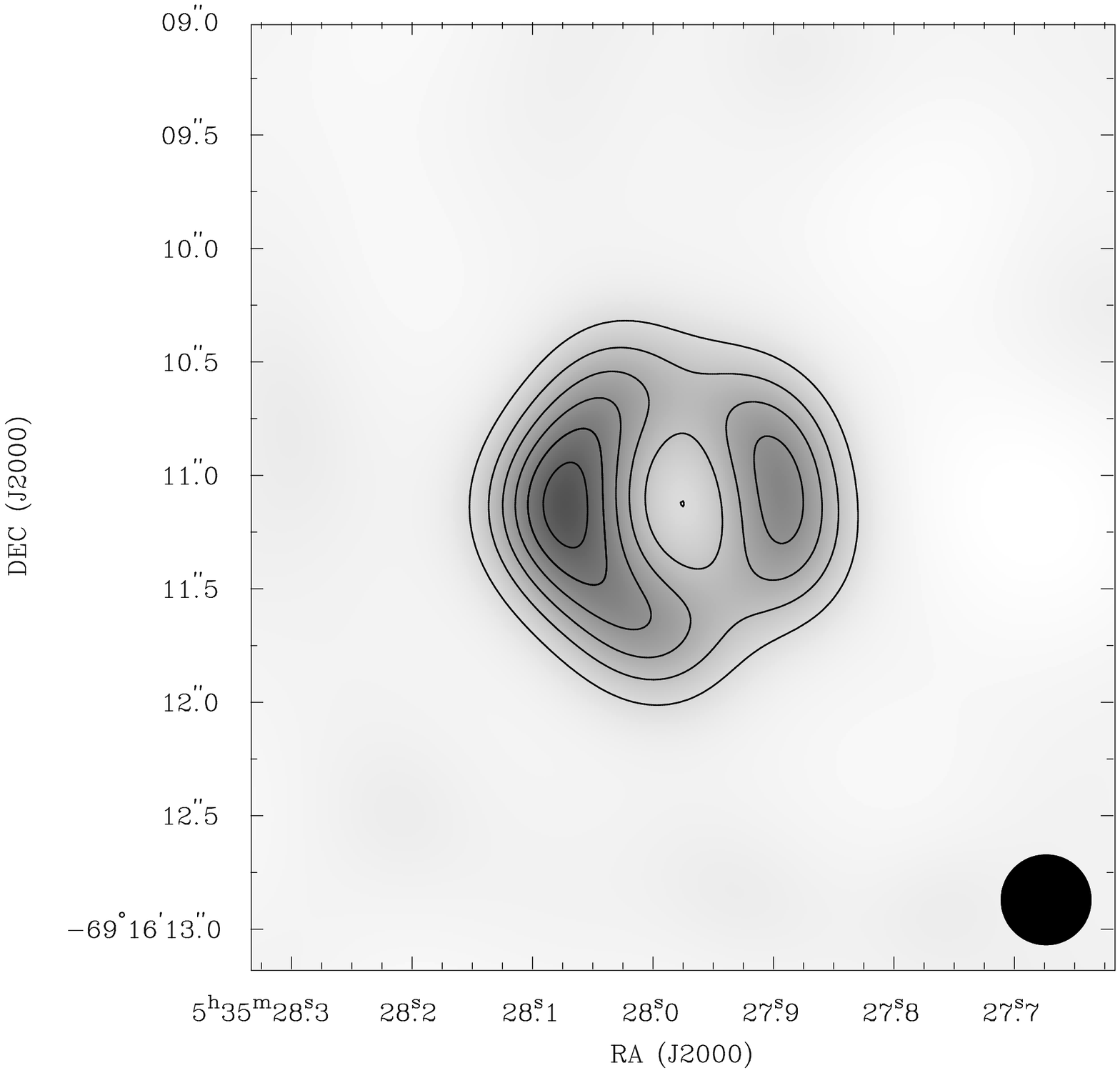,height=10.0cm,angle=270}}
\caption{}
\label{1994_sn}
\end{figure}

\begin{figure}
\centerline{\psfig{file=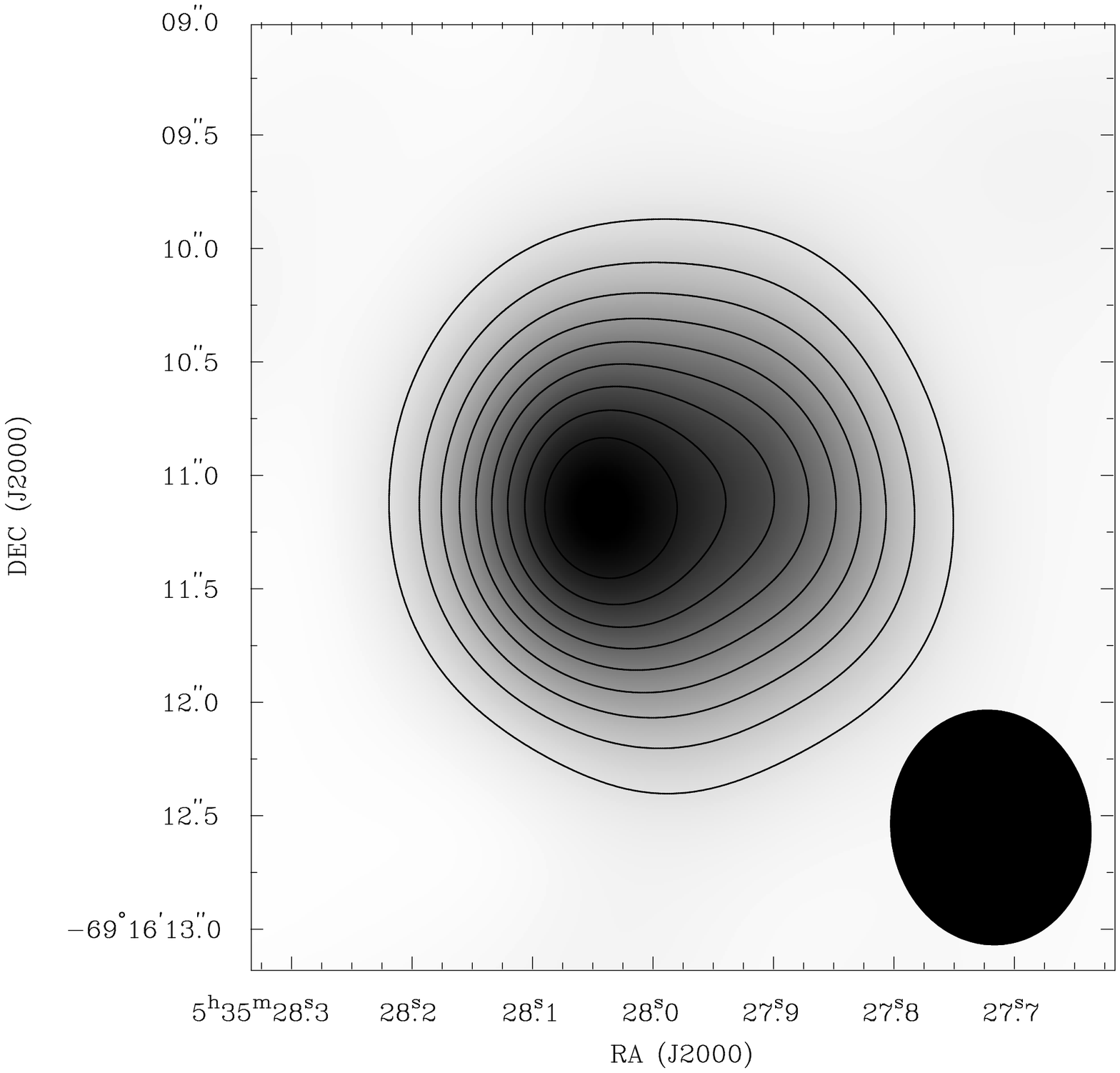,height=10.0cm,angle=270}}
\vspace{2cm}
\centerline{\psfig{file=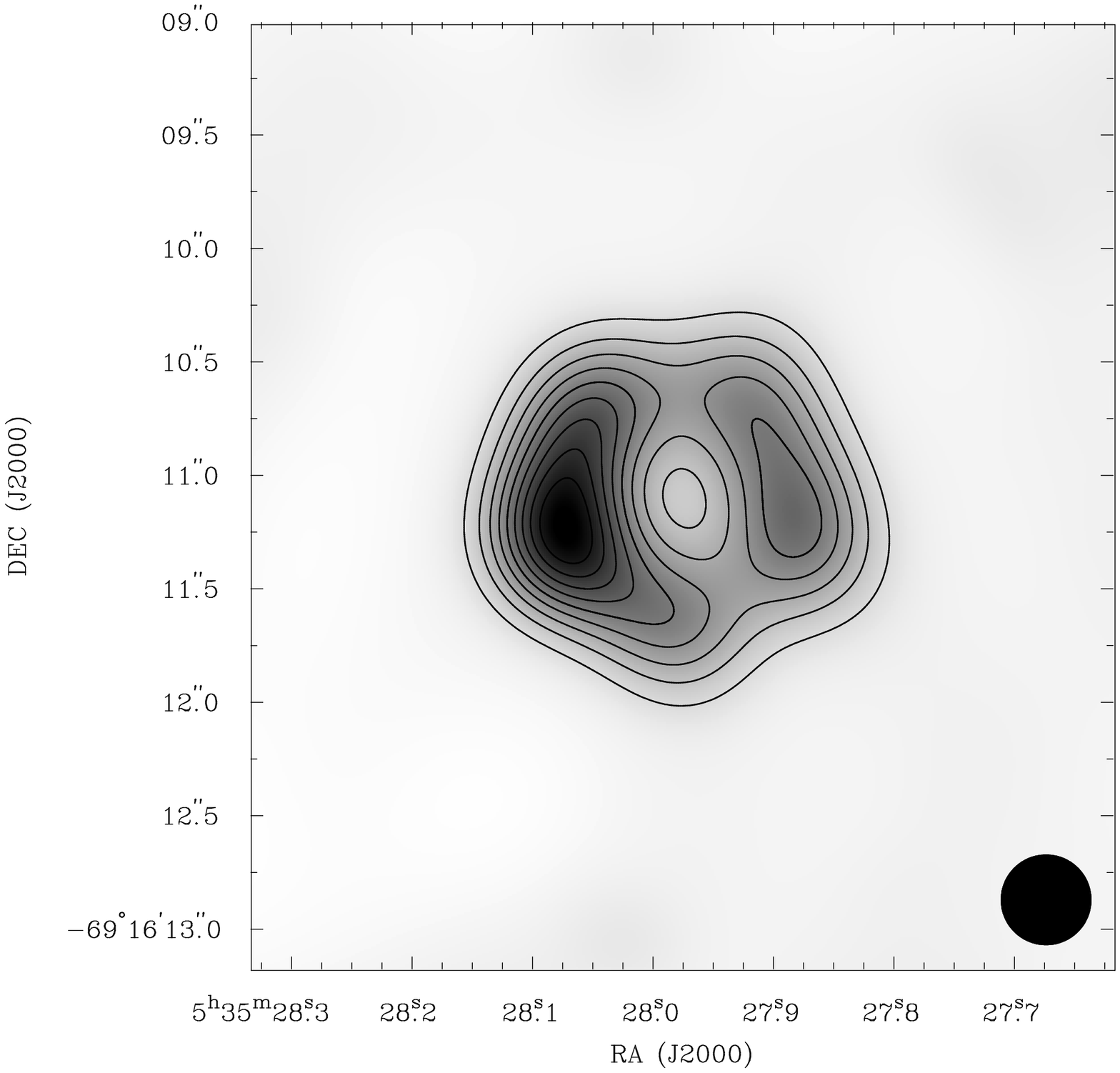,height=10.0cm,angle=270}}
\caption{}
\label{1995_sn}
\end{figure}

\begin{figure}
\psfig{file=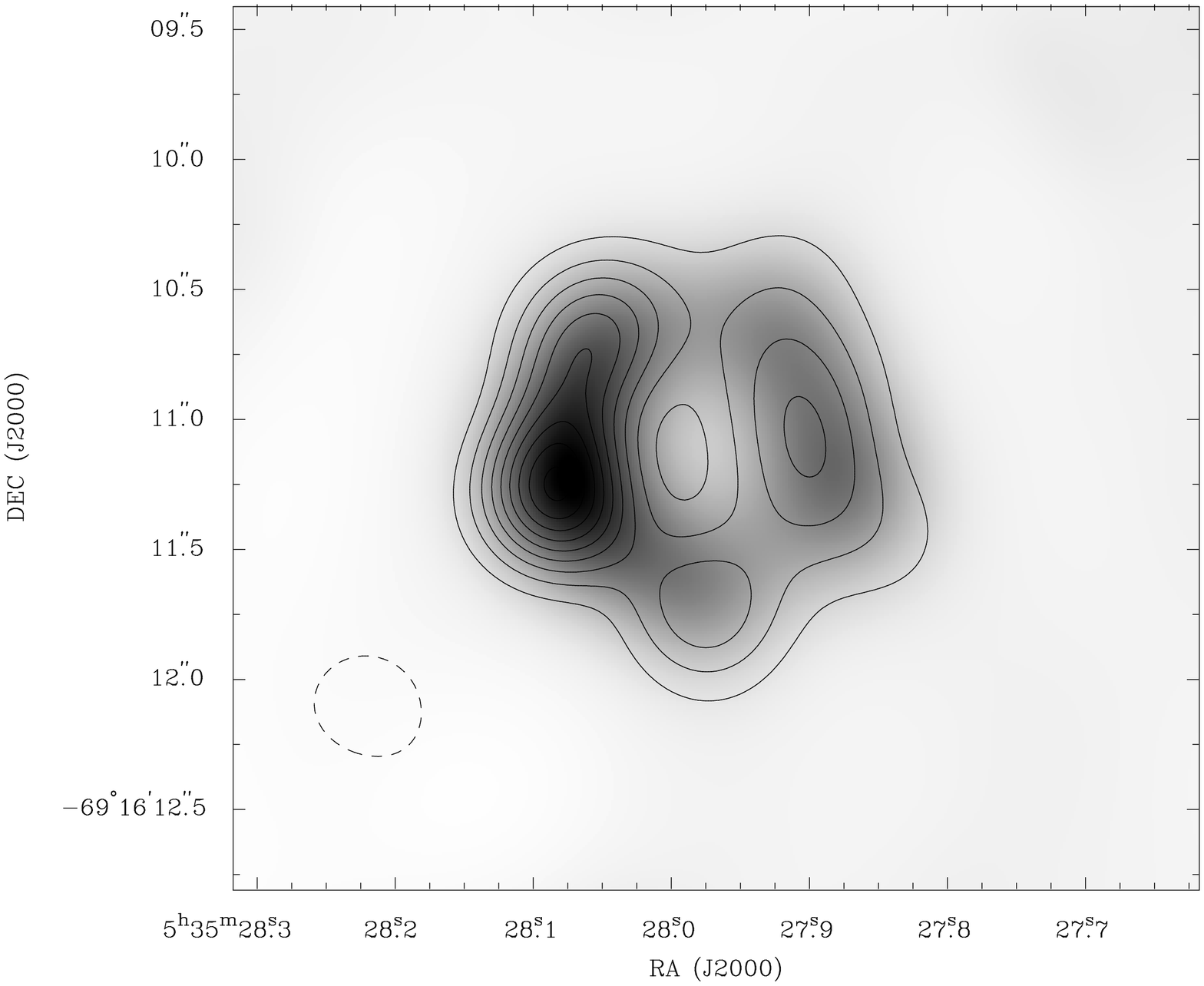,height=10.0cm,angle=270}
\caption{}
\label{1995-1993_sn}
\end{figure}

\begin{figure}
\centerline{\psfig{file=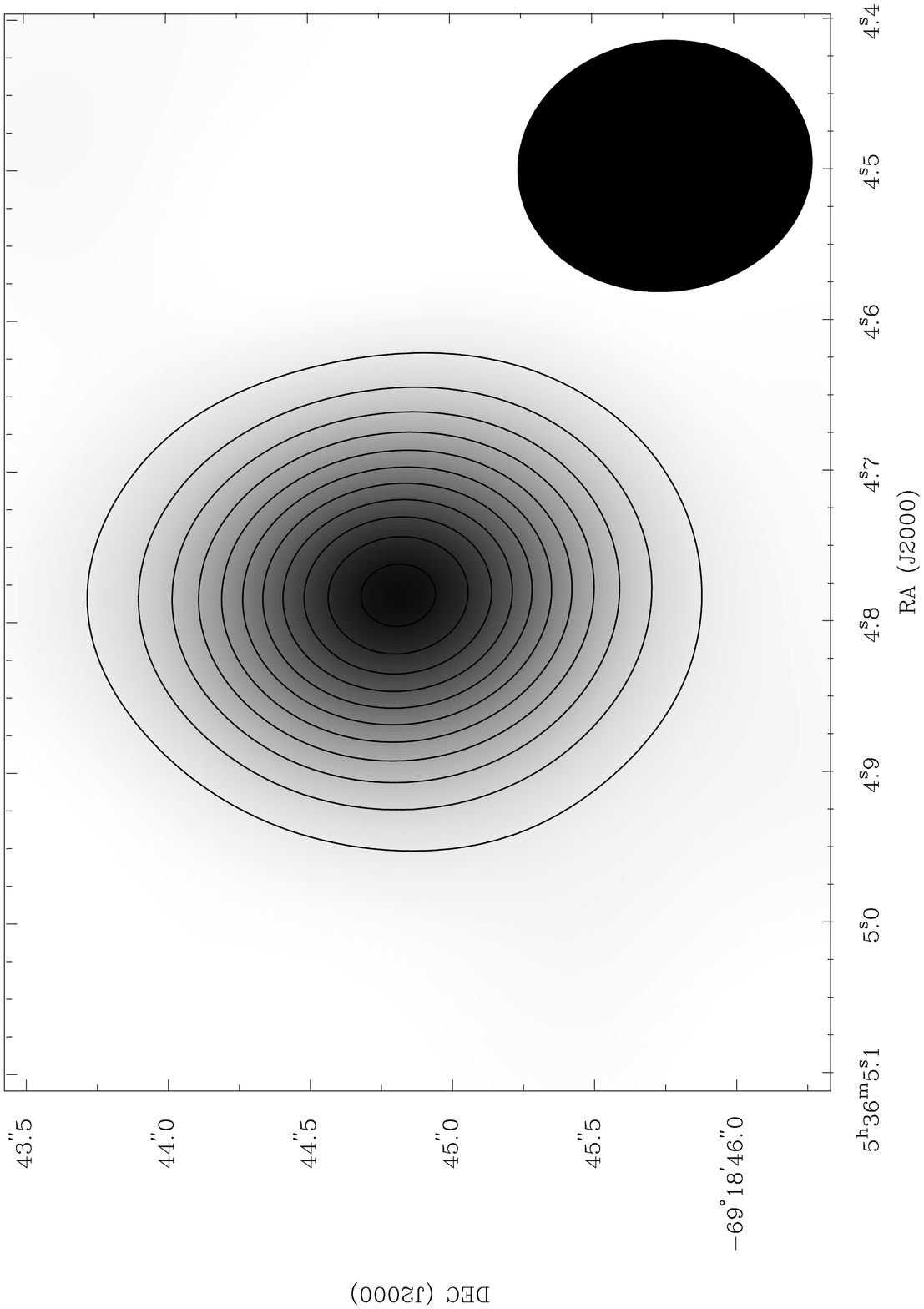,height=10.0cm,angle=270}}
\vspace{2cm}
\centerline{\psfig{file=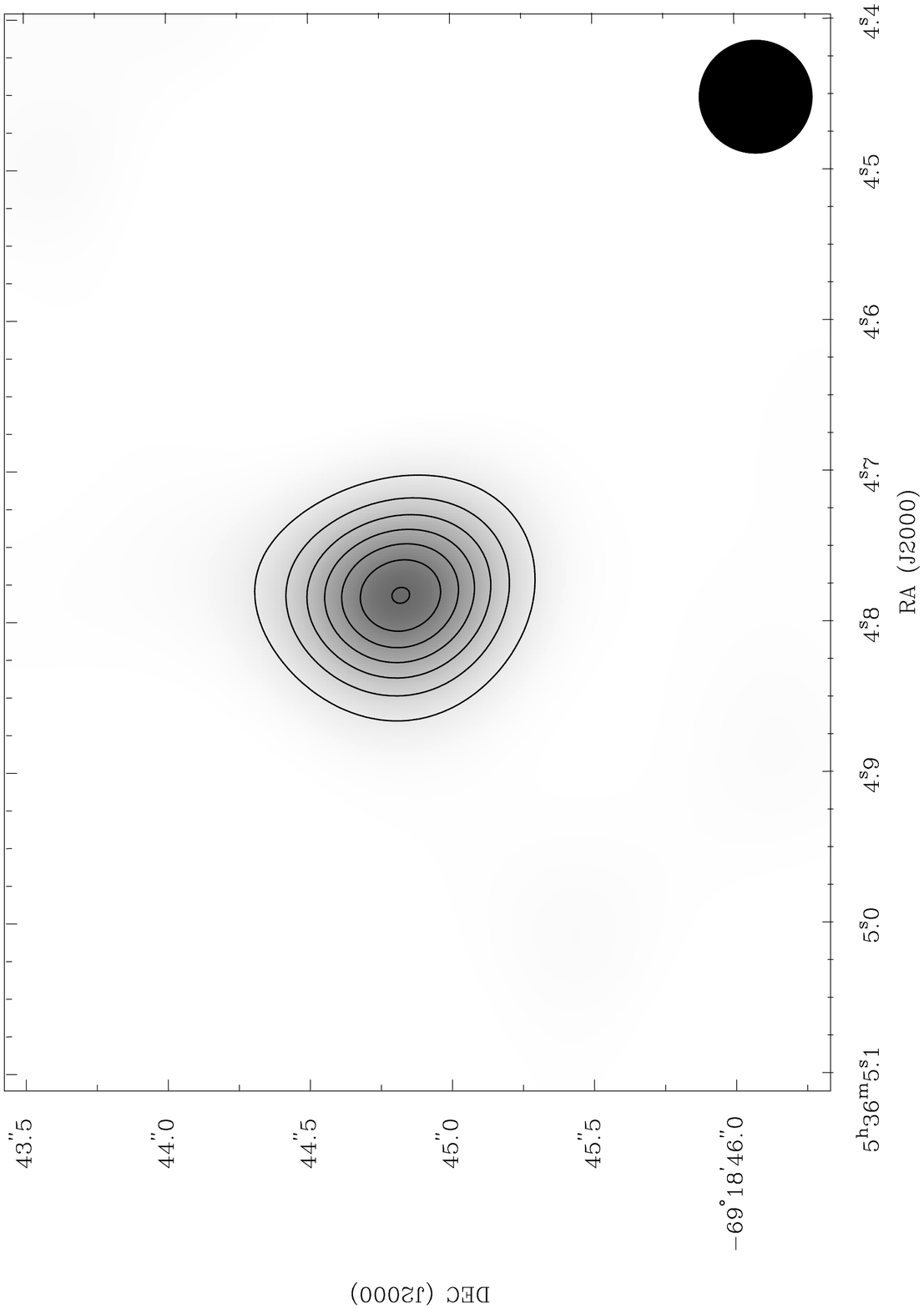,height=10.0cm,angle=270}}
\caption{}
\label{1995_src1}
\end{figure}

\begin{figure}
\centerline{\psfig{file=hst_overlay.ps,height=20.0cm}}
\caption{}
\label{hst_overlay}
\end{figure}

\begin{figure}
\centerline{\psfig{file=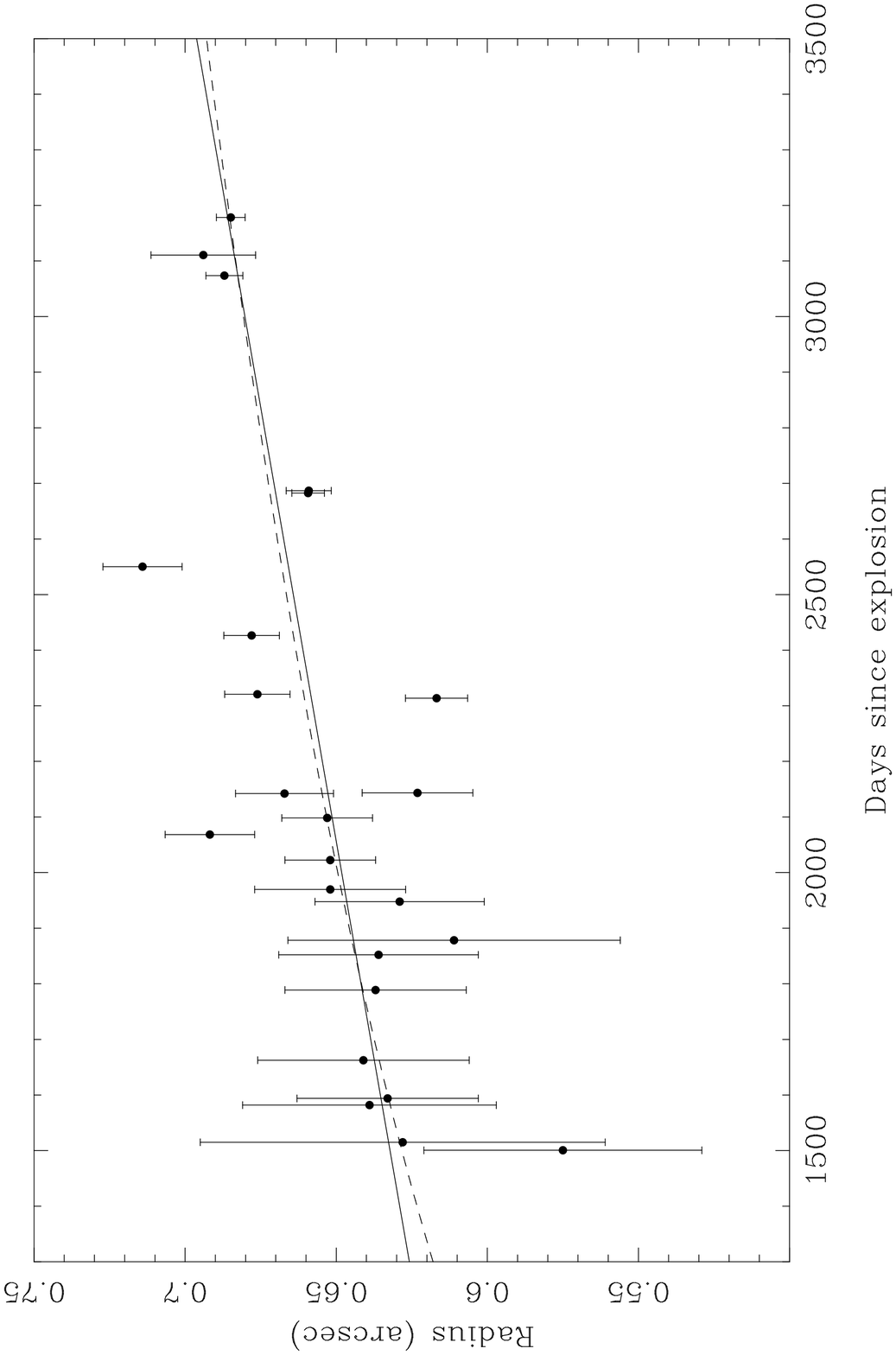,height=20.0cm}}
\caption{}
\label{radius_plot}
\end{figure}

\end{document}